# Laboratory development of a heterodyne interferometric system for translation and tilt measurement of the proof mass in the space gravitational wave detection

Xin Xu and Yidong Tan*
State Key Laboratory of Precision Measurement Technology and Instruments, Department of Precision Instruments, Tsinghua University, Beijing 100084, China
Tanyd@mail.tsinghua.edu.cn

**ABSTRACT**

Laser heterodyne interferometry plays a key role in the proof mass's monitor and control by measuring its multiple degrees of freedom motions in the Space Gravitational Wave Detection. Laboratory development of polarization-multiplexing heterodyne interferometer (PMHI) using quadrant photodetectors (QPD) is presented in this paper, intended for measuring the translation and tilt of a proof mass. The system is of symmetric design, which can expand to five degrees of freedom measurements based on polarization-multiplexing and differential wavefront sensing (DWS). The ground-simulated experimental results demonstrate that a measurement noise of 3 pm/Hz$^{1/2}$ and 2 nrad/Hz$^{1/2}$ at 1 Hz have been achieved respectively. The tilt-to-length error is dominated by geometric misalignment for the current system, the coupling of which is at micrometer level within a tilt range of ±500 μrad.

## 1. INTRODUCTION

Recently, several scientific missions aim to put the heterodyne interferometers into space, such as LISA, Taiji and Tianjin, opening a new window to the subhertz gravitational-wave universe [1-4]. In these missions, the core technology is to construct an optical readout system for the relative distance monitor between two freely proof masses, which is also known as the gravitational reference sensor. To successfully detect the gravitational waves, the measurement sensitivity of the optical readout system needs to reach ~1 pm/Hz$^{1/2}$ within the frequency band from 0.1 mHz to 1 Hz [5]. Such high precision level in the long baseline demands that the translation and tilt of the proof mass should be monitored and pre-controlled down within the noise requirement (about 1 pm/Hz$^{1/2}$ of translation and 1 nrad/Hz$^{1/2}$ of tilt at the frequency band of 1mHz-1Hz). Optical readout signals also help to refer and control the position of the proof mass and the beam pointing.

In 2009, Thilo Schuldt et al designed a ground-simulated system, which verified the feasibility of using the laser heterodyne interferometers for picometer displacement measurement. The translation measurement is 2 pm/Hz$^{1/2}$ noise level and the tilt measurement is 1 nrad/Hz$^{1/2}$ noise level, both for frequency above 0.1 Hz [6]. Intensity stabilization and phase-lock control technology are also implemented to achieve such high sensitivity. Another research of simultaneous measurement of translation and tilt has been reported through a dual-heterodyne laser interferometer in 2015, with a sensitivity noise level of 1 pm/Hz$^{1/2}$ and 0.4 nrad/Hz$^{1/2}$ at 1 Hz [7].

LISA Pathfinder, launched by European Space Agency (ESA) in 2015, is seen as an important milestone in space gravitational wave detection development [8]. It contains four interferometers to monitor the relative translation and tilt between two proof masses at a distance of ~30cm. The space-tested results are impressive, as the optical readout sensitivity reaches up to the requirement of the LISA mission. The newest test results show that a measurement noise of 0.032 pm/ Hz$^{1/2}$ has been successfully achieved [9]. Inspired by the great success of LISA Pathfinder, many research groups continue to develop high-precision and multiple degrees of freedom heterodyne interferometers.

Ziren Luo, et al constructed an on-ground laser interferometer prototype platform in the past decade [10-13]. The path-length measurement sensitivity of the ground-simulated test reached 5 pm/ Hz$^{1/2}$ in 2020 [13]. This noise level meets the requirement of the Taiji Pathfinder mission, providing an optical readout solution to the position and control of the freely proof mass.

A multi-axis heterodyne interferometry system has been constructed and tested as a short-arm measurement demonstration for LISA in the paper [14]. The results show that the translation measurement sensitivity of the prototype is better than 10 pm/Hz$^{1/2}$ for frequencies above 4 mHz and below 1 pm/Hz$^{1/2}$ for frequencies above 35 mHz, and the tilt measurement sensitivity is below 10 nrad/Hz$^{1/2}$ for frequencies above 4 mHz and below 1 nrad/Hz$^{1/2}$ for frequencies above 100 mHz.

In 2022, the working group from Tianqin proposed a heterodyne interferometric system based on differential wavefront sensing, which can measure six degrees of freedom of the proof mass [15]. This all-optical sensing system theoretically provides a higher precision of the translation and tilt measurement than the optical-capacitive combined sensing method in LISA Pathfinder. Moreover, the capability to monitor six degrees of freedom at the same time can obtain more useful information on the motion of the freely falling proof mass.

Recent developments of laser heterodyne interferometers have heightened the need for lower measurement noise and multiple degrees of freedom measurement. In this paper, we construct a polarizing heterodyne interferometric system with two quadrant photodetectors employed. Due to the highly symmetric topology of the optical layout, the measurement noise caused by vibration, temperature drift, and laser frequency fluctuations can be effectively suppressed. Furthermore, when the AOM drivers are used as the reference sources, the system with polarization multiplexing and two QPDs can extend to five degrees of freedom measurement. The paper is organized into four main sections. First, the optical design and working principle are introduced in Part 2. Then, the experimental interferometric system is constructed and introduced in Part 3, with which the performance results of stability, resolution, range and tilt-to-length noise are shown in Part 4. The conclusion and the future arrangements are presented in Part 5.

## 2 Method

Fig. 1 shows the optical configuration of the proposed polarizing heterodyne interferometric system. It contains two interferometers, each for the three degrees of freedom measurement using two quadrant photodetectors. Two beams ($f_1$ and $f_2$) with 45° linear polarization are incident and split by the polarizing beam splitter (PBS$_{1,2}$). The detailed paths of the measurement beam and the reference beam for the two interferometers are listed in Table 1. Taking the second interferometer as an example, shown in Fig. 2, the reference beam of horizontal linear polarization is incident onto the quadrant photodetector (QPD$_2$) after a polarizing beam splitter (PBS$_2$) and a polarizer (P$_2$). The measurement beam of the second interferometer passes through the polarizing beam splitter (PBS$_1$), and the transmissive part is reflected by the reflector and the proof mass. Twice passing through the quadrant wave-plate changes the polarizing direction of the backward beam, so this beam can be reflected by the polarizing beam splitter (PBS$_1$). Then, the measurement beam is combined by the polarizing beam splitter (PBS$_2$) with the reference beam. The polarizer (P$_2$) keeps the same polarizing part and the beat signals are detected by the quadrant photodetector (QPD₂) with the active area divided into the quadrants A-B-C-D.

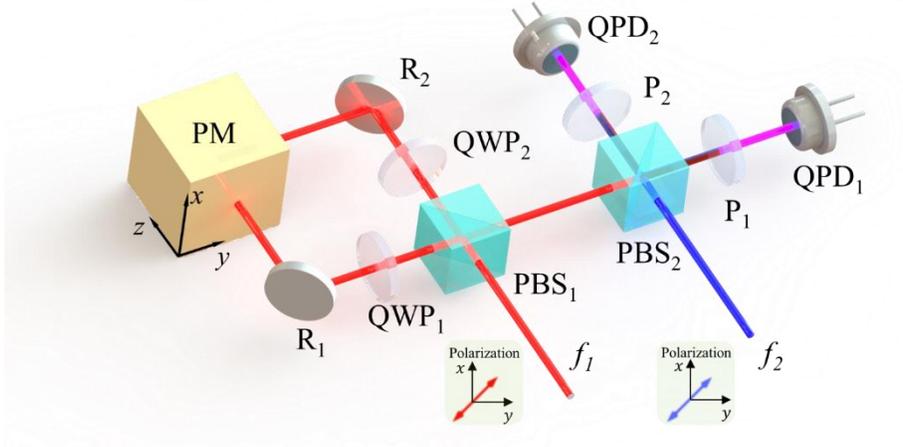

Fig. 1 Schematic diagram of the optical configuration. (Polarization Multiplexing Heterodyne Interferometer, PMHI)

| Path1_m | $f_1$→(45°)PBS₁→(↕)QWP₁→(+/-)R₁→(+/-)TM→(+/-)R₁→(+/-)QWP₁→(↔)PBS₁→(↔)PBS₂→(↔)P₁→(45°)QPD₁ | QPD₁ |
|---|---|---|
| Path1_r | $f_2$→(45°)PBS₂ → (↕)P₁ → (45°)QPD₁ | $\Delta z\ \theta x\ \theta y$ |
| Path2_m | $f_1$→(45°)PBS₁→(↔)QWP₂→(+/-)R₂→(+/-)TM→(+/-)R₂ →(+/-)QWP₂→(↕)PBS₁→(↕)PBS₂→(↕)P₂→(45°)QPD₂ | QPD₂ |
| Path2_r | $f_2$→(45°)PBS₂→(↔)P₂→ (45°)QPD₂ | $\Delta x\ \theta y\ \theta z$ |

↔: horizontal linear polarization; ↕: vertical linear polarization; 45°: 45° linear polarization; +/-: circular polarization;

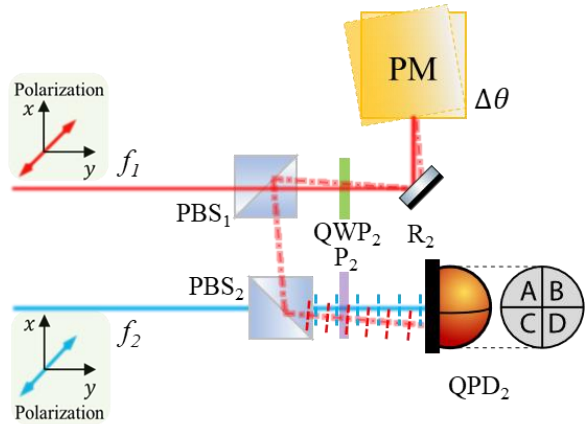

Fig. 2 Schematic diagram of three degrees of freedom measurement using one quadrant photodetector.

Based on differential wavefront sensing (DWS) and longitudinal pathlength sensing (LPS), a quadrant photodetector can measure three degrees of freedom [15-16]. In the proposed system, there are two interferometers, each having a quadrant photodetector. Therefore, five degrees of freedom can be measured, as the tilt along the *x*-axis is common for the two QPDs. The measurement can be expressed as:

$$\Delta x = \frac{\lambda}{4\pi} \frac{\phi_{1A} + \phi_{1B} + \phi_{1C} + \phi_{1D}}{4}$$

$$\Delta z = \frac{\lambda}{4\pi} \frac{\phi_{2A} + \phi_{2B} + \phi_{2C} + \phi_{2D}}{4}$$

$$\theta x \approx \frac{\lambda}{4\sqrt{2\pi}D} \cdot (\phi_{1A} - \phi_{1B} + \phi_{1C} - \phi_{1D}) \approx \frac{\lambda}{4\sqrt{2\pi}D} \cdot (\phi_{2A} - \phi_{2B} + \phi_{2C} - \phi_{2D})$$

$$\theta y \approx \frac{\lambda}{4\sqrt{2\pi}D} \cdot (\phi_{1A} + \phi_{1B} - \phi_{1C} - \phi_{1D})$$

$$\theta z \approx \frac{\lambda}{4\sqrt{2\pi}D} \cdot (\phi_{2A} + \phi_{2B} - \phi_{2C} - \phi_{2D})$$

where $\phi_{ij}$ ($i = 1,2$ $j = A, B, C, D$) is the quadrant phase change calculated from the interference wavefront detected by the two QPDs. $\Delta x$ and $\Delta z$ represent the translations of the proof mass, $\theta x$, $\theta y$ and $\theta z$ represent the three tilt degrees of freedom respectively. $\lambda$ is the measured laser wavelength and $D$ is the beam diameter on the photodetector. $\lambda/4\sqrt{2\pi}D$ is the first-order coefficient factor of the tilt measurement, the typical value of which is approximately $10^{-3} \sim 10^{-4}$ rad/rad [15].

## 3 Experimental setup

The experimental system contains three main parts, including laser source unit, the optical bench of five degrees of freedom interferometers and the phase readout subsystem. The basic principles of the heterodyne frequency generation and the phase readout are similar to many former researches [6, 7, 15]. In this paper, we use a non-planar ring-oscillator (NPRO) type Nd:YAG laser (Mephisto-S 1064) providing ~500 mW of output beam power. Part of it after the isolator is coupled into a polarization-preserving fiber and injected into the optical bench by a fiber collimator (Thorlabs, CFP5-1064A). The coupled beam is split equally via a splitter (Thorlabs, BS014), and its frequency is shifted by two acoustic-optical modulators (Gooch&Housego, AOMO 3080-197). After passing a half wave-plate and a polarizer, we obtain two beams of $f_1$ and $f_2$ with 45° linear polarization. The frequency difference of the two drivers is set at 1 MHz in our system, and it provides a reference signal for the phase readout subsystem.

Fig. 3 shows the optical bench, which is covered by an acrylic box to decrease the air perturbation. The power of the input beams is approximately 5 mW and its waist diameter is about 1 mm. The optical bench is fixed on a marble table with an independent ground foundation. The beat signals are acquired by the photodetectors and sent to the phase demodulation subsystem. The subsystem consists of two lock-in amplifiers (Zurich Instrument, HF2LI), which can be synchronized for four-channel phase measurement. It should be noted that five degrees of freedom measurement need an eight-channel phasemeter, which is currently under development in our laboratory. Therefore, we can only measure three degrees of freedom at one time. In the experiments, all-optical components are carefully positioned to achieve equal-arm interference signals, therefore reducing the common mode noise of laser frequency fluctuations and ambient disturbance noise. In the stability experiments, the reflector acts as the target in that it can be tightly fixed on the bench to decrease the vibration. In the resolution and range experiments, the target is a quartz cube with six gold-plated surfaces, which is placed on a six-axis nanopositioning stage (Physik Instruments, P-562.6CD, resolution 1 nm and 100 nrad, range 200 μm and 1000 μrad).

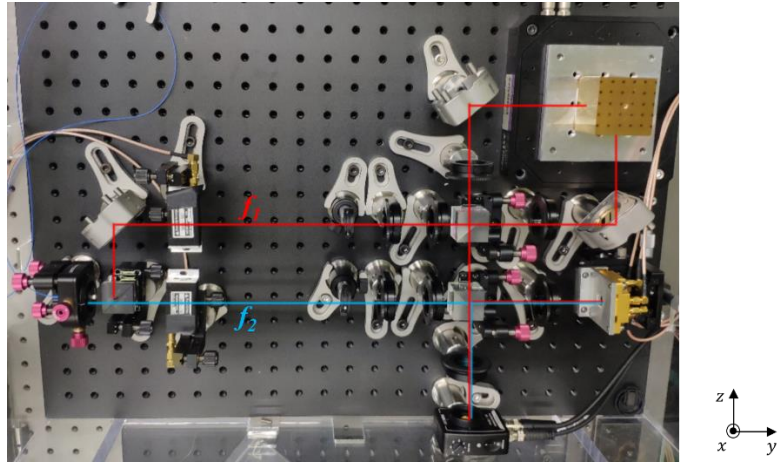
Fig. 3 Test results of the sensitivity baseline of the interferometer system.

## 4 Results

**Sensitivity baseline**

We first test the fundamental noise level of the optical readout system. This helps to figure out the sensitivity baseline and is necessary for the following translation and tilt measurements. We divide the beat-frequency signal from one photodetector into two parts and trace the phase difference between them. The demodulator band is set at 20 Hz and the sampling rate is 225 Sa/s. The amplitude spectrum density (ASD) and linear amplitude spectrum density (LASD) analysis of the test data is carried out using a Matlab toolbox, which is developed by the Max Planck Institute for Gravitational Physics [17]. The curves in Fig. 4 show that the electronic readout and phase measurement noise of the interferometric system is lower than $4\pi \times 10^{-6}$ rad/Hz$^{1/2}$ in the frequency band of 1mHz-1Hz. This proves that the system can meet up the signal demodulation requirement of picometer translation and nanoradian tilt measurement. Note that the optical path of the proposed heterodyne interferometer has a round trip, thus a 2pi change in phase corresponds to a half-wavelength translation.

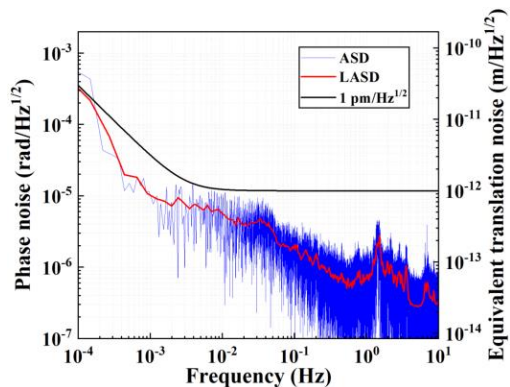
Fig. 4 Test results of the sensitivity baseline of the interferometer system.

**Stability**

Then we evaluate the measurement noise of five degrees of freedom. To well test the limited sensitivity, we use the reflector fixed directly on the bench as the measurement target and the reference is from the beat signal of another polarization path. The ASD curves of the translation and tilt are plotted in Fig. 5. It shows that the measurement noise can reach 3 pm/ Hz$^{1/2}$ and 2 nrad/ Hz$^{1/2}$ at 1 Hz. Learning from many previous researches on the measurement noise of heterodyne interferometers, we make a preliminary conclusion of the possible noise causes in the different frequency bands. For the current optical system, the noise below 1 mHz is mainly caused by the temperature variation, and it can be corrected by a precise temperature control during the test or a detrend data processing of the test results. The noise above 1 Hz is from the vibration of the test bench, which can be sharply reduced by a differential detection of common-path measurement. At the frequency band of 1 mHz – 1 Hz the noise increases with a 1/$f$-behaviour, which is owing to the optical path difference and the laser frequency fluctuation. Nevertheless, the above analysis is a preliminary analysis of the possible noise resources, and the specific reasons and a quantitative analysis need a further study of the constructed interferometers.

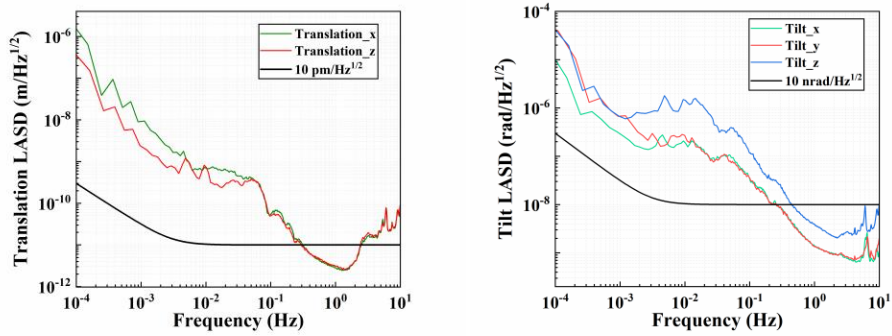

Fig. 5 Measurement results of five degrees of freedom. Left: ASD of translation $\Delta x$ and $\Delta z$; Right: ASD of tilt $\theta x$, $\theta y$ and $\theta z$.

**Resolution**

In order to test the minimum incremental motion of the five degrees of freedom measurement, we use a commercial nanopositioning stage as an actuation stage. The resolution of the translation is 5 nm and of the tilt is 100 nrad, as shown in Fig. 6.

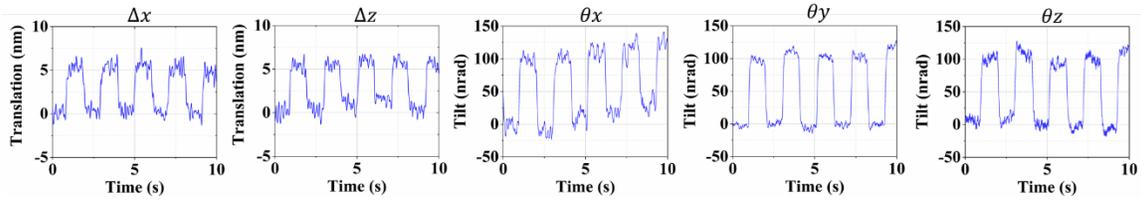

Fig. 6 Resolution results of five degrees of freedom.

For the translation resolution experiments, the effective resolution of the used actuation stage is 1 nm, but the constructed interferometer cannot distinguish such tiny displacement. To figure out this point, a short-term drift of the target with different installation methods is tested. We find that the stability of the target fixed on the stage is about 3 nm, while the stability of the target fixed directly on the bench is about 0.4 nm, as shown in Fig. 7. Therefore, it is reasonable that the constructed interferometer has a resolution better than 5 nm, as the current stage does not have enough stability to carry out such tiny displacement resolution experiments. If we use the standard deviation of a short-term drift as the resolution, the minimum incremental motion of the translation is approximately 46 pm.

For the tilt resolution experiments, the effective resolution of the used actuation stage is 100 nrad, and the constructed interferometer can clearly distinguish such tilt. Like the analysis of the translation, we also test the short-term tilt drift with different installation conditions. Owing to the differential wavefront sensing, the vibration can be well eliminated, as the sensing beams are passing through almost the same path. The drift is assumed to be affected by the instant temperature fluctuations. The minimum incremental motion of the tilt is about 10 nrad calculated from the standard deviation of a 10s tilt drift.

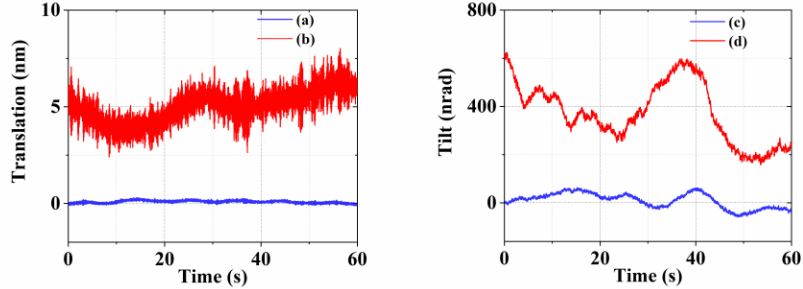

Fig. 7 Measurement results of five degrees of freedom. (a)Translation measurement when the target is fixed directly on the bench (b)Translation measurement when the target is fixed on the nanopositioning stage (c)Tilt measurement when the target is fixed directly on the bench. (d)Tilt measurement when the target is fixed on the nanopositioning stage.

### Range

Apart from the stability and the resolution, the range of five degrees of freedom interferometers is also tested. A back-and-forth full-range motion of translation and tilt are launched out by the nanopositioning stage, and the tested results are plotted in Fig. 8.

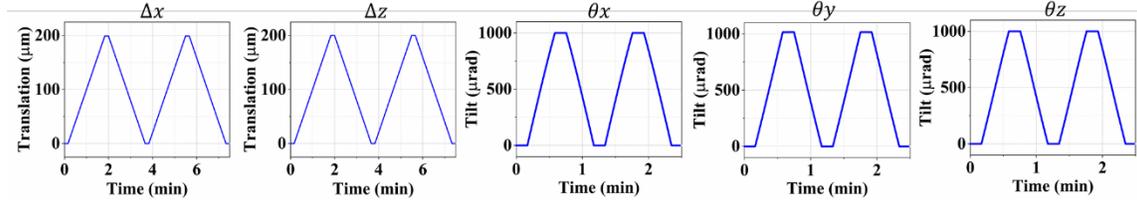

Fig. 8 Measurement results of five degrees of freedom.

### Tilt-to-length coupling

Tilt-to-length (TTL) coupling is a main noise source in space gravitational wave detection. In ideal conditions, the translation and the tilt should not affect each other. However, when the proof mass suffers a tilt jitter, it will cause a translation motion, coupling into the longitudinal readout. This coupling error would seriously ruin the measurement sensitivity of the optical system [18-19]. In the preliminary experiments, we monitor the translation change when the proof mass tilts in a range of 1000 μrad driven by the nanopositioning stage (P-562.6CD), as shown in Fig. 9.

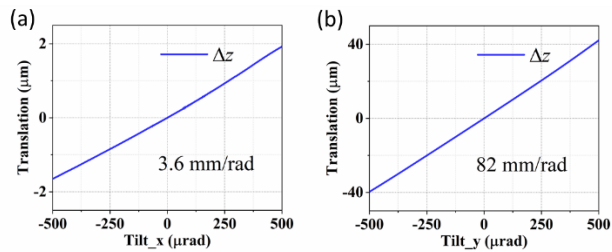

Fig. 9 Tilt-to-length error measurement results of the proof mass. (a) $\theta x$-$z$: the translation results of z-direction when the proof mass rotates around the x-axis. (b) $\theta y$-$z$: the translation results of z-direction when the proof mass rotates around the y axis.

The motion coordinate system is not coincident with the proof mass's coordinate system when we use a nanopositioning stage as the motion generator. The height of the beam from the rotating axis y is measured as ~80 mm, as shown in Fig. 4. When the stage rotates, it will make the proof mass move in the z-direction. The error is calculated to be ~80 μm (80 mm×tan(1mrad)), which is consistent with the measured result in Fig. 8(b). For the tilt-to-length error of $\theta x$-$z$, we put the proof mass in the center of the nanopositioning stage, and the distance from the rotating axis x locates in the micrometer level. Therefore, we believe that the tilt-to-length (TTL) coupling is mainly caused by the geometric misalignment for the current interferometric system.

The current TTL coupling is about 3.6 mm/rad around the tilt_x axis and 82 mm/rad around the tilt_y axis. It remains a gap comparing the current measurement performance of TTL noise with the requirement of 25 μm/rad in LISA [19]. Nevertheless, the developed optical system of multiple degrees of freedom measurement would be a fine experimental platform for the further study of tilt-to-length coupling of heterodyne interferometers.

In the future step, we are to use a tip/tilt platform with a reflector as the target, rather than a nanopositioning stage, to avoid the misalignment between the motion coordinate system and the proof mass's coordinate system. Apart from the geometric misalignment, some nongeometric effects will also cause the tilt-to-length coupling, such as the beam-axis offset and the wavefront distortion. These mechanisms are more complicated and need also to be carefully suppressed.

In existing reports, LISA designs a double-lens imaging system to suppress the TTL noise, which has reached the requirements of inter-satellite detection sensitivity [19]. The solution to the problem is to suppress the off-axis degree of the Gaussian beam. Nevertheless, the rotation center of the tilt jitter is not necessarily on the mass, but may be at any spatial position in the payload satellite. Therefore, the tilt jitter may still affect the final measurement sensitivity. To figure out this point, we are to construct a heterodyne interferometric system, as its optical layout is shown in Fig. 10.

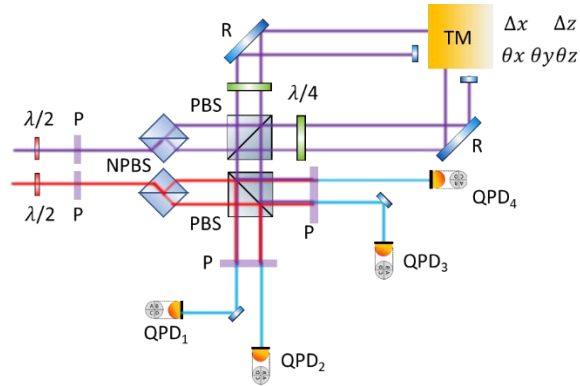

Fig. 10 Schematic of the optical layout for multiple degrees of freedom measurement of the proof mass, based on polarization multiplexing, dual-beam interferometry and differential wavefront sensing.

Its design idea has drawn on many previous works, including dual-beam interferometry [6-7] and differential wavefront sensing [16]. In this system, the translation readout can be acquired in two methods. One is from the polarization-multiplexing heterodyne interferometer (PMHI), which is similar to the proposed interferometer above. Another way is to use the dual-beam heterodyne interferometer (DBHI), in which the beat from QPD1 is the reference signal and the beat from QPD2 is the measurement signal. Similarly, two-dimensional translation can be measured by polarization multiplexing, and the tilts of the target can be measured by differential wavefront sensing [16] or dual-beam interferometry [7].

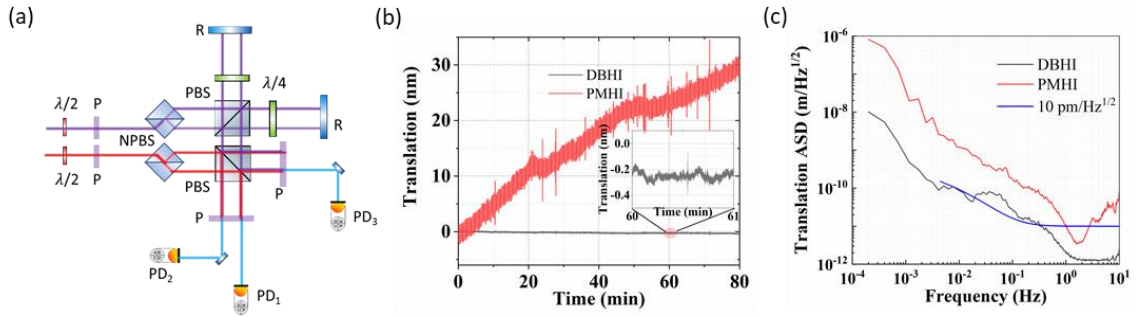

Fig. 11 Preliminary test of the heterodyne interferometric system with a combination of PMHI and DBHI. (a) test setup (b) translation stability of time-domain. (c) translation stability of frequency-domain.

The translation measurement stability of the PMHI and the DBHI is tested. The optical setup and results of time-domain and frequency-domain are shown in Fig. 11. DBHI readout has a sensitivity improvement than the PMHI, as the optical path of the former design is common and the target is the same reflector. The initial results prove that the translation measurement noise is lower than 2 pm/ in the frequency band 1-10Hz.

## 5 Conclusion

In conclusion, we propose and construct an optical heterodyne interferometric system of translation and tilt measurement. The optical design can be extended for five degrees of freedom measurement based on polarization multiplexing and differential wavefront sensing. The common-path and symmetric design help to reach a low measurement noise level. The stability, resolution, sensing range of the translation and tilt are tested. The translation measurement noise is 3 pm/ $Hz^{1/2}$ at 1 Hz and the tilt measurement noise is 2 nrad/$Hz^{1/2}$ at 1 Hz. The laboratory development of such optical heterodyne interferometers for the multiple-dimension motion measurement may act as a high-precision reference sensor and use it for the control of the freely falling proof mass in the space gravitational wave detection.

We also launch a preliminary study of the tilt-to-length coupling for the constructed optical interferometric system. The tilt-to-length (TTL) coupling is believed to be caused by the geometric misalignment for the current interferometric system. The current TTL coupling is about 3.6 mm/rad around the tilt_x axis and 82 mm/rad around the tilt_y axis.

In the next stage, we plan to put the developed system in a vacuum chamber. To suppress the measurement noise, the vibration should be passively isolated and the operating temperature needs to keep constant. Moreover, the TTL noise from geometric effects and nongeometric effects needs to be further studied and carefully suppressed. We believe that the test environment control and the TTL noise suppression can effectively improve the sensitivity of the translation and tilt measurement for the proof mass. It still needs a lot to accomplish, and we hope this laboratory development of heterodyne interferometric system could provide a potential solution to the translation and tilt measurement of the freely falling proof mass in the future space gravitational wave detection.